\documentclass[allclo]{FBSart}
\usepackage{amsfonts}
\usepackage{amsmath}
\usepackage{amssymb}
\usepackage{multirow}
\usepackage{bm,graphicx}
\usepackage{wrapfig}

\title{ Analysis of three-nucleon forces effects in the $A=3$ system }

\author{A. Kievsky\instnr{1}}
\instlist{
Istituto Nazionale di Fisica Nucleare, Largo B. Pontecorvo 3, 56127 Pisa, Italy}

\runningauthor{A. Kievsky}
\runningtitle{Analysis of the three-nucleon force in the $A=3$ systems }

\newcommand {\be}{\begin{equation}}
\newcommand {\ee}{\end{equation}}
\newcommand {\bea}{\begin{eqnarray}}
\newcommand {\ea}{\end{eqnarray*}}
\newcommand {\ba}{\begin{eqnarray*}}
\newcommand {\eea}{\end{eqnarray}}

\begin{document}

\maketitle

\begin{abstract}
Using modern nucleon-nucleon interactions in the description of the
$A=3,4$ nuclear systems the $\chi^2$ per datum results to be much bigger
than one. In particular it is not possible to reproduce the three- 
and four-nucleon binding energies and the $n-d$ scattering length
simultaneously. This is one manifestation of 
the necessity of including a three-nucleon force in the nuclear Hamiltonian. 
In this paper we perform an analysis of some, widely used, 
three-nucleon force models. We analyze their 
capability to describe the aforementioned quantities and,
to improve their description, 
we propose modifications in the parametrization of the models.
The effects of these new parametrization are studied in
some polarization observables at low energies.
\end{abstract}

\maketitle

\section{Introduction}
Realistic nucleon-nucleon (NN) potentials reproduce the 
experimental NN scattering data up to energies
of $350$ MeV with a $\chi^2$ per datum close to one. 
However, the use of these potentials in the description 
of the three- and four-nucleon bound and scattering states gives a
$\chi^2$ per datum much bigger than one (see for example Ref.~\cite{kiev01}).
In order to reproduce correctly the three-nucleon bound state energy,
different three-nucleon force (TNF) models have been introduced 
as the Tucson-Melbourne (TM), Brazil (BR) and the 
Urbana IX (URIX) models \cite{tm,brazil,urbana}. These models are based
on the exchange mechanism of two pions between three nucleons. 
In the case of the TM model, it has been revisited
within a chiral symmetry approach~\cite{friar:a},
and it has been demonstrated that the contact term present in it
should be dropped. This new TM potential, known as TM', has been
subsequently readjusted~\cite{tmp} and the final operatorial structure
coincides with that one given in the TNF of Brazil.
TNF models based on $\pi\rho$ and $\rho\rho$ meson exchange
mechanisms have also been derived~\cite{coon} and their effects have been studied in
the triton binding energy~\cite{stadler}.
More recently, TNFs have been derived~\cite{epelbaum02}
using a chiral effective field theory at next-to-next-to-leading order. 
A local version of these interactions (hereafter referred as N2LOL) 
can be found in Ref.~\cite{N2LO}. At next-to-next-to-leading order, the TNF has
two unknown constants that have to be determined. It is a common
practice to determine these parameters from the three- and four-nucleon binding 
energies ($B$($^3$H) and $B$($^4$He), respectively). 
It should be noticed that in this procedure, the three- and four- nucleon
systems are described in the framework of the non relativistic quantum mechanics.
Relativistic corrections to the few-nucleon binding energies have been
studied in Ref.~\cite{pieper01} and, recently, the three-nucleon Faddeev equations
have been solved in a Poincar\'e invariant model~\cite{witala05}. These efforts are
directed to establish if some of the discrepancies observed between experimental
data and theoretical descriptions, as for example the
minimum of the $N-d$ differential cross section, can be reduced if relativistic
corrections are taken into account.

The $n-d$ doublet scattering length $^2a_{nd}$ 
is correlated, to some extent, to the $A=3$ binding
energy through the so-called Phillips line~\cite{phillips,bedaque}. 
However the presence of TNFs could break this correlation. 
Therefore $^2a_{nd}$ can be used as an independent
observable to evaluate the capability of the interaction models to
describe the low energy region. 
In Ref.~\cite{report} results for different combinations of NN interactions
plus TNF models are given. We report the results for the quantities of
interest in Table I. From the table, we can observe that
the models are not able to describe simultaneously the $A=3,4$
binding energies and $^2a_{nd}$. In Ref.~\cite{kiev10}
a comparative study of the aforementioned TNF models has been performed.
The AV18~\cite{av18} was used as the reference NN interaction
and the three-nucleon interaction models were added to it. 
Different parametrizations of the URIX, TM' and
N2LOL TNF have been constructed in order
to reproduce, in conjunction with the AV18 interaction, $B$($^3$H),
$B$($^4$He) and $^2a_{nd}$. In a second step some polarization observables 
in $p-d$ scattering at $E_{lab}=3$ MeV have been studied. 
In the case of the vector analyzing powers, it was observed that
the predictions of the different parametrizations appear in narrow
bands with different positions for each model. Compared to the
original AV18+URIX model, the results obtained using
the parametrizations of the N2LOL model
were slightly better, in particular for $A_y$ and $iT_{11}$.
The results obtained using the parametrizations of the TM' 
were of the same quality. Conversely, the proposed parametrizations 
for the URIX model produced a much worse description of
$A_y$ and $iT_{11}$ than the original URIX model. 
A possible explanation for this fact is the
particular behavior of the profile functions $Y(r)$ and $T(r)$
used in the construction of the model. To this end, in the
present paper, we study a different functional form of the
profile functions $Y(r)$ and $T(r)$.

\section{Three Nucleon Force Models}

In Ref.~\cite{report} the description of bound states and zero-energy
states for $A=3,4$ has been reviewed in the context of the HH method.
In Table~\ref{tb:table1} we report results for
the triton and $^4$He binding energies as well as for the doublet
$n-d$ scattering length $^2a_{nd}$ using the
AV18 and the N3LO-Idaho~\cite{entem} NN potentials and using the following
combinations of two- and three-nucleon interactions:
AV18+URIX, AV18+TM' and N3LO-Idaho+N2LOL.
The results are compared to the experimental values of the binding
energies and $^2a_{nd}$~\cite{doublet}.

From the table we observe that the results obtained 
using an interaction model that includes a TNF are close to the
corresponding experimental values. In the case of the AV18+TM', the
strength of the TM' potential has been fixed to reproduce the
$^4$He binding energy and the triton binding energy is 
slightly underpredicted. Conversely,
the strength of the URIX potential
has been fixed to reproduce the triton binding energy, giving too much
binding for $^4$He. The strength of the N2LOL potential has been
fixed to reproduce simultaneously the triton and the $^4$He binding
energies. In the three cases the predictions for the doublet scattering 
length are not in agreement with the experimental value.

\begin{table}[h]
\caption{The triton and $^4$He binding energies $B$ (MeV),
and  doublet scattering length $^2a_{nd}$ (fm)
calculated using the AV18 and the N3LO-Idaho
two-nucleon potentials, and
the AV18+URIX, AV18+TM' and N3LO-Idaho+N2LOL two- and three-nucleon interactions.
The experimental values are given in the last row.}
\label{tb:table1}
\begin{tabular}{@{}llll}
\hline
Potential & $B$($^3$H) & $B$($^4$He) & $^2a_{nd}$ \cr
\hline
AV18            & 7.624    & 24.22   & 1.258 \cr
N3LO-Idaho      & 7.854    & 25.38   & 1.100 \cr
AV18+TM'        & 8.440    & 28.31   & 0.623 \cr
AV18+URIX       & 8.479    & 28.48   & 0.578 \cr
N3LO-Idaho+N2LOL & 8.474    & 28.37   & 0.675 \cr
\hline
Exp.            & 8.482    & 28.30   & 0.645$\pm$0.003$\pm$0.007 \cr
\hline
\end{tabular}
\end{table}

Following Ref.~\cite{kiev10} we give a brief description of the TNF models. 
Starting from the general form
\begin{equation}
W= \sum_{i<j<k} W(i,j,k)  \;\; ,
\label{eq:wijk}
\end{equation}
a generic term can be decomposed as
\begin{equation}
W(1,2,3)= aW_a(1,2,3)+bW_b(1,2,3)+dW_d(1,2,3)+c_DW_D(1,2,3)+c_EW_E(1,2,3) \; .
\label{eq:w123}
\end{equation}
Each term corresponds to a different mechanism and has a different operatorial 
structure.  The specific form of these three terms
in configuration space is the following:
\begin{equation}
\begin{aligned}
& W_a(1,2,3) = W_0(\bm\tau_1\cdot\bm\tau_2)(\bm\sigma_1\cdot \bm r_{31})
           (\bm\sigma_2\cdot \bm r_{23}) y(r_{31})y(r_{23}) \\
& W_b(1,2,3)= W_0 (\bm\tau_1\cdot\bm\tau_2) [(\bm\sigma_1\cdot\bm\sigma_2)
  y(r_{31})y(r_{23})  \\
 &\hspace{2cm} + (\bm\sigma_1\cdot \bm r_{31})
    (\bm\sigma_2\cdot \bm r_{23})(\bm r_{31}\cdot \bm r_{23})
  t(r_{31})t(r_{23}) \\
 & \hspace{2cm} + (\bm\sigma_1\cdot \bm r_{31})(\bm\sigma_2\cdot \bm r_{31})
  t(r_{31})y(r_{23}) \\
 &\hspace{2cm} + (\bm\sigma_1\cdot \bm r_{23})(\bm\sigma_2\cdot \bm r_{23})
  y(r_{31})t(r_{23})] \\
& W_d(1,2,3)=W_0(\bm\tau_3\cdot\bm\tau_1\times\bm\tau_2)  
   [(\bm\sigma_3\cdot \bm\sigma_2\times\bm\sigma_1)y(r_{31})y(r_{23}) \\
  & \hspace{2cm}+ (\bm\sigma_1\cdot \bm r_{31})
    (\bm\sigma_2\cdot \bm r_{23})(\bm\sigma_3\cdot\bm r_{31}\times \bm r_{23})
  t(r_{31})t(r_{23}) \\
 & \hspace{2cm} 
  + (\bm\sigma_1\cdot \bm r_{31})(\bm\sigma_2\cdot \bm r_{31}\times\bm\sigma_3)
  t(r_{31})y(r_{23}) \\
 &\hspace{2cm} + (\bm\sigma_2\cdot \bm r_{23})(\bm\sigma_3\cdot \bm r_{23}\times
  \bm\sigma_1) y(r_{31})t(r_{23})]\;\; ,
\end{aligned}
\end{equation}
with $W_0$ an overall strength.
The $b$- and $d$-terms are present in the three models whereas the $a$-term
is present in the TM' and N2LOL and not in URIX. Here we are
interested in the profile functions $y(r)$ and $t(r)$.
In the first two models these functions are obtained from the following 
function
\begin{equation}
f_0(r)=\frac{12\pi}{m_\pi^3}\frac{1}{2\pi^2}
 \int_0^\infty dq q^2 \frac{j_0(qr)}{q^2+m_\pi^2}F_\Lambda(q) 
\label{eq:f0r}
\end{equation}
where $m_\pi$ is the pion mass and
\begin{equation}
\begin{aligned}
&y(r)=\frac{1}{r} f^\prime_0(r)   \\
&t(r)=\frac{1}{r} y^\prime(r)  \,\,\ .
\end{aligned}
\label{eq:y0r}
\end{equation}
The cutoff function $F_\Lambda$ 
in the TM' or Brazil models is taken as 
$F_\Lambda=[(\Lambda^2-m_\pi^2)/(\Lambda^2+q^2)]^2$. 
In the N2LOL model it is taken as $F_\Lambda=\exp(-q^4/\Lambda^4)$. 
The momentum cutoff $\Lambda$ is a parameter of the model
fixing the scale of the problem in momentum space.
In the N2LOL, it has been fixed to $\Lambda=500$ MeV, whereas in the TM' model the ratio
$\Lambda/m_\pi$ has been varied to describe the triton or $^4$He binding energy
at fixed values of the constants $a$,$b$ and $d$. In the literature the TM' potential
has been used many times with typical 
values around $\Lambda= 5\; m_\pi$.

In the URIX model the radial dependence of the $b$- and $d$-terms is given in terms
of the functions 
\begin{equation}
\begin{aligned}
&Y(r)={\rm e}^{-x}/x\,\xi_Y   \\
&T(r)=(1+3/x+3/x^2)Y(r)\,\xi_T
\end{aligned}
\label{eq:Y0r}
\end{equation}
with $x=m_\pi r$ and the cutoff functions are defined as 
$\xi_Y=\xi_T=(1-{\rm e}^{-cr^2})$, with $c=2.1$ fm$^{-2}$. 
This regularization has been used in the AV18 potential
as well. Since the URIX model has been constructed in
conjunction with the AV18 potential, the use of the same
regularization was a choice of consistency.
The relation between the functions $Y(r),T(r)$
and those of the previous models is:
\begin{equation}
\begin{aligned}
& Y(r)=y(r)+T(r)  \\
& T(r)=\frac{r^2}{3}t(r)\,\,\, . 
\end{aligned}
\label{eq:T0r}
\end{equation}
With the definition given in Eq.(\ref{eq:f0r}), the asymptotic behavior of
the functions $f_0(r)$, $y(r)$ and $t(r)$ is:
\begin{equation}
\begin{aligned}
&f_0(r\rightarrow\infty)\rightarrow \frac{3}{m_\pi^2}\frac{{\rm e}^{-x}}{x} \\
&y(r\rightarrow\infty)\rightarrow -\frac{3{\rm e}^{-x}}{x^2}
       \left(1+\frac{1}{x}\right)  \\
&t(r\rightarrow\infty)\rightarrow \frac{3}{r^2}\frac{{\rm e}^{-x}}{x}
       \left(1+\frac{3}{x}+\frac{3}{x^2}\right) \;\; .
\end{aligned}
\label{eq:f0rasymp}
\end{equation}
To be noticed that with the normalization chosen for $f_0$, the functions $Y$ and
$T$ defined from $y$ and $t$ and those ones defined in the URIX model 
coincide at large separation distances. 

The last two terms in Eq.(\ref{eq:w123}) correspond to a two-nucleon (2N) contact term
with a pion emitted or absorbed ($D$-term) and to a three-nucleon (3N)
 contact interaction ($E$-term). Their local form, in configuration space,
derived in Ref.~\cite{N2LO}, is
\begin{equation}
\begin{aligned}
& W_D(1,2,3)= W_0^D (\bm\tau_1\cdot\bm\tau_2) \{(\bm\sigma_1\cdot\bm\sigma_2)
  [y(r_{31})Z_0(r_{23})+Z_0(r_{31})y(r_{23})]  \\
 & \hspace{2cm} + (\bm\sigma_1\cdot \bm r_{31})(\bm\sigma_2\cdot \bm r_{31})
  t(r_{31})Z_0(r_{23}) \\
 &\hspace{2cm} + (\bm\sigma_1\cdot \bm r_{23})(\bm\sigma_2\cdot \bm r_{23})
  Z_0(r_{31})t(r_{23})\} \\
& W_E(1,2,3) = W_0^E(\bm\tau_1\cdot\bm\tau_2) Z_0(r_{31})Z_0(r_{23})  \,\, .
\end{aligned}
\end{equation}
The constants $W_0^D$ and $W_0^E$ fix the strength of these terms.
In the case of the URIX model the $D$-term is absent whereas 
the $E$-term is present without the isospin operatorial
structure and it has been included as purely phenomenological, without
justifying its form from a particular exchange mechanism. Its radial dependence
has been taken as $Z_0(r)=T^2(r)$.
In the N2LOL model, the function $Z_0(r)$ is defined as
\begin{equation}
Z_0(r)=\frac{12\pi}{m_\pi^3}\frac{1}{2\pi^2}
 \int_0^\infty dq q^2 j_0(qr) F_\Lambda(q)
\label{eq:z0r}
\end{equation}
with the same cutoff function used before, $F_\Lambda(q)=\exp(-q^4/\Lambda^4)$.
In the TM' model the $D$- and $E$-terms are absent.

\begin{figure}[t]
\vspace{1cm}
\includegraphics[scale=0.6,angle=0]{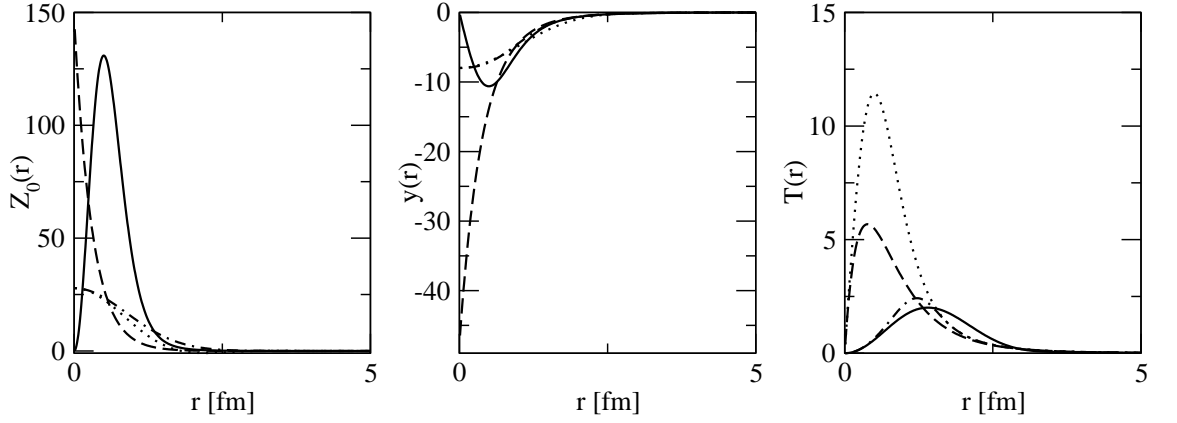}
\caption{ The $Z_0(r)$, $y(r)$ and $T(r)$ functions as functions of the
interparticle distance $r$ for the URIX (solid line), TM' (dashed line) and
N2LOL (dotted line) models. The dotte-dashed line shows the one-parameter
functions defined in Eqs.(~\protect\ref{eq:z0rp}) and (~\protect\ref{eq:y0rp}).}
\label{fig:functions}
\end{figure}

In order to analyze the different short range structure of the TNF models,
in Fig.~\ref{fig:functions} we compare the dimensionless functions 
$Z_0(r)$, $y(r)$ and $T(r)$ for the three models under consideration.
In the TM' model using the definition
of Eq.(\ref{eq:z0r}) and using the corresponding cutoff function we can define:

\begin{equation}
 Z^{TM}_0(r)=\frac{12\pi}{m_\pi^3}\frac{1}{2\pi^2}
 \int_0^\infty dq q^2 j_0(qr) \left(\frac{\Lambda^2-m_\pi^2}{\Lambda^2+q^2}
\right)^2 
 = \frac{3}{2}\left(\frac{m_\pi}{\Lambda}\right)
   \left(\frac{\Lambda^2}{m_\pi^2}-1\right)^2 {\rm e}^{-\Lambda r}  \;\; .
\label{eq:z0rtm}
\end{equation}
This function is shown in the first panel of Fig.~\ref{fig:functions}
as a dashed line.
From the figure we can see that, in the case of the URIX model, the functions
$Z_0(r)$ and $y(r)$ go to zero as $r\rightarrow 0$. This is not the case for the
other two models and is a consequence of the choice to regularize the $Y$
and $T$ functions adopted in the URIX. The function $Z^{TM}_0(r)$ has been
introduced in Ref.~\cite{kiev10} to add a repulsive term to the TM' model.
In fact, it was shown that without it, the AV18+TM' model was unable to
reproduce simultaneously the triton binding energy and the doublet scattering 
length for reasonable values of the TM' strength parameters.

\section{Parametrization of the profile functions $Z_0(r)$, $y(r)$ and $T(r)$}

In this section we study a possible parametrization of
the profile functions $Z_0(r)$, $y(r)$ and $T(r)$.
The function $Z_0$ is defined in Eq.(\ref{eq:z0r}), its behavior for
small values of $r$ can be derived from the expansion of the Bessel
function as
\begin{equation}
Z_0(r\rightarrow 0)=\frac{12\pi}{m_\pi^3}\frac{1}{2\pi^2}
 \int_0^\infty dq q^2 \left[ 1-\frac{q^2r^2}{6}+\ldots \right]F_\Lambda(q)
\label{eq:z0r1}
\end{equation}
For a sharp cutoff this integral can be approximated as
\begin{equation}
Z_0(r\rightarrow 0)\approx\frac{12}{m_\pi^3}\frac{1}{2\pi}
 \left[ \frac{\Lambda^3}{3}-r^2\frac{\Lambda^5}{30}+\ldots \right]
\approx\frac{2}{\pi}\left(\frac{\Lambda^3}{m_\pi^3}\right)
\left[ 1-\frac{r^2\Lambda^2}{10}\right]
\label{eq:z0r2}
\end{equation}
Therefore in the following we propose the one-parameter form
of the function
\begin{equation}
Z_0(r)=
Z_0(0){\rm e}^{-r^2\Lambda^2/10}
\label{eq:z0rp}
\end{equation}
where $Z_0(0)$ can be taken from Eq.(\ref{eq:z0r}) using a particular 
form of cutoff. In the case of sharp cutoff it results:
$Z_0(0)=\frac{2}{\pi}\left(\frac{\Lambda^3}{m_\pi^3}\right)$.

The functions $y(r)$ and $t(r)$ are defined in Eq.(\ref{eq:y0r}).
Their short range behavior, after expanding the corresponding Bessel function,
are
\begin{equation}
\begin{aligned}
 y(r\rightarrow 0)&=-\frac{12\pi}{m_\pi^3}\frac{1}{2\pi^2}\frac{1}{r}
 \int_0^\infty dq q^3 \left[ \frac{qr}{3}(1-\frac{q^2r^2}{10}+\ldots) \right]
  \frac{F_\Lambda(q)}{q^2+m_\pi^2} \\
&=y(0)+\frac{1}{2}r^2 t(0)+\ldots  \\
 t(r\rightarrow 0)&=-\frac{12\pi}{m_\pi^3}\frac{1}{2\pi^2}\frac{1}{r^2}
 \int_0^\infty dq q^4 \left[ \frac{q^2r^2}{15}(1-\frac{q^2r^2}{14}+\ldots) \right]
  \frac{F_\Lambda(q)}{q^2+m_\pi^2} \\
&=t(0)-\frac{1}{2}r^2 t_2+\ldots 
\end{aligned}
\label{eq:y0r1}
\end{equation}
where we have introduced the corresponding values at $r=0$ and the quantity
$t_2$ in the second term of $t(r)$.
In Eq.(\ref{eq:f0rasymp}) the asymptotic behavior of the functions $y(r)$ and
$t(r)$ are given. Recalling that $3T(r)=r^2 t(r)$ and considering the short-range 
behavior indicated
above, we will analyze the following one-parameter 
$r$-space form of the functions $y$ and $T$
\begin{equation}
\begin{aligned}
&y(r)=-\frac{3{\rm e}^{-x}}{x^2} \left(1+\frac{1}{x}\right)
       \left(1-{\rm e}^{-x^3|y(0)|/3}\right) \\
&T(r)=\;\; \frac{{\rm e}^{-x}}{x}
       \left(1+\frac{3}{x}+\frac{3}{x^2}\right)
       \left(1-{\rm e}^{-x^3r^2|t(0)|/9}\right)
\end{aligned}
\label{eq:y0rp}
\end{equation}
These functions are shown in Fig.~\ref{fig:functions} with the dot-dashed
line. They have been calculated using the cutoff of the N2LOL model.
As expected they are close to the profile functions of the N2LOL
potential. Calculations in the $A=3$ systems using 
these profile $r$-space functions are analyzed in the next section.

\section{Results in the $A=3$ system}

In the previous section we have presented the one-parameter profile functions,
$Z_0(r)$, $y(r)$ and $T(r)$,
obtained from a regularization of their asymptotic form. The 
regularization was perfomed in order to match the short range behavior
of the profile functions defined in the N2LOL potential. As it is shown
explicitly in Fig.~\ref{fig:functions}, these functions are very different
from those defined in the Urbana potential in which a different regularization
was used. In order to study the sensitivity in the description of the
$A=3$ system to different profile functions, we construct a modification of the
Urbana model in which the profile functions defined in 
Eqs.(\ref{eq:z0rp}) and (\ref{eq:y0rp}) are used in place of the original ones.
The parameters $Z_0(0)$, $y(0)$ and $t(0)$ are calculated 
using the cutoff of the N2LOL model, $F_\Lambda=\exp(-q^4/\Lambda^4)$, with
$\Lambda=500$ MeV. As mentioned, in the URIX model, only the
$b$-, $d$ and $E$-terms are included. The corresponding strengths are
fixed by the constants $A^{PW}_{2\pi}$, $D^{PW}_{2\pi}$ and $A_R$. Their
original values are shown in the first row of Table~\ref{tb:urbe}. 
Changing the form of the profile functions the values of the constants
has to be fixed. Three sets of values, selected to reproduce the triton
binding energy and $^2a_{nd}$, are given in Table~\ref{tb:urbe} together
with some mean values calculated from the triton wave function.

\begin{table}[h]
\caption{Mean values of the triton kinetic energy, the two-nucleon
potential energy $V(2N)$, and the attractive, $V_A(3N)$, and repulsive,
$V_R(3N)$, contributions of the TNF to the triton binding energy
using the AV18+URIX potential for the specified values of the parameters.
In the last column $^2a_{nd}$ is given.}
\label{tb:urbe}
\begin{tabular}{@{}cccccccc}
\hline
$A^{PW}_{2\pi}$& $D^{PW}_{2\pi}$& $A_R$ & $T$ & $V(2N)$ & $V_A(3N)$ & $V_R(3N)$ &
  $^2a_{nd}$ \cr
  [MeV] &  & [MeV] & [MeV] & [MeV] & [MeV] & [MeV] & [fm]  \cr
\hline
-0.0293 & 0.25   & 0.0048  & 51.259 & -58.606 & -1.126 & 1.000 & 0.578 \cr
-0.1200 & 0.25   & 0.0108  & 50.110 & -57.360 & -1.747 & 0.525 & 0.643 \cr
-0.1200 & 0.50   & 0.0155  & 50.193 & -57.328 & -2.097 & 0.759 & 0.645 \cr
-0.1200 & 0.75   & 0.0229  & 50.331 & -57.211 & -2.735 & 1.143 & 0.644 \cr
\hline
\end{tabular}
\end{table}

\begin{figure}[htb]
\includegraphics[width=14cm,height=16cm,angle=0]{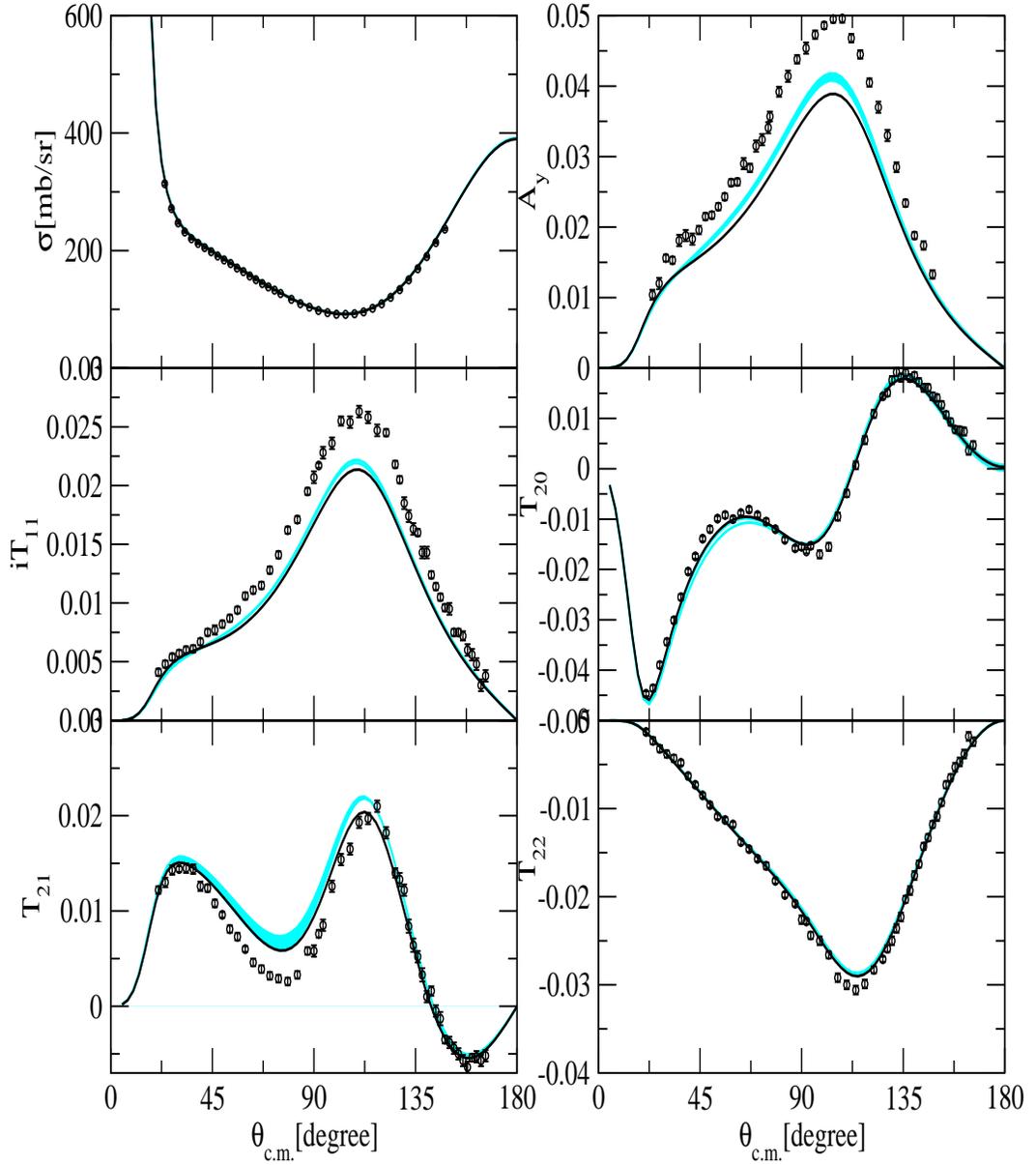}
\caption{ (Color on line)
Differential cross section and vector and tensor polarization
observables at $E_{lab}=3$ MeV using the AV18+URIX model with the parameters
given in Table~\ref{tb:urbe} (cyan band). The predictions of the original
AV18+URIX model, given in the first row of the table, are shown as a solid line.
The experimental points from Ref.~\protect\cite{shimizu} are also shown.}
\label{fig:fig2}
\end{figure}

From the table we can observe that with the proposed parametrization the
attractive part of the TNF is bigger and, due to the fact that
the profile functions are smoother, there is 
a reduction of the mean value of the kinetic energy.
In order to extend further the analysis, $p-d$ scattering observables at 
$E_{lab}=3$ MeV have been calculated using the three sets of constants
and the new form of the profile functions. The results for the differential
cros section, the vector analyzing powers $A_y$ and $iT_{11}$ and the
tensor analyzing powers $T_{20}$, $T_{21}$ and $T_{22}$, are shown in
Fig.\ref{fig:fig2} and compared to the predictions of the original 
URIX model and the experimental data. From the figure we can observe that,
besides a small improvement in $A_y$ and $iT_{11}$, all the models 
describe the data with similar quality. However only the models with
the profile functions of Eqs.(\ref{eq:z0rp}) and (\ref{eq:y0rp})
reproduce the experimental value of $^2a_{nd}$. The fact that the vector
analyzing powers improve very little with the new parametrization can be
taken as a further evidence that the spin-isospin structure of the URIX
is incomplete and different forms could in principle be included~\cite{kievsky99}.

\section{Conclusions}

Due to the fact that some of the widely used TNF models do not
reproduce simultaneously the triton and $^4$He binding energies and
the $n-d$ doublet scattering length, possible modifications of their 
parametrizations have been analyzed. To this end we
have used the AV18 as the reference NN interaction and we have analyze
possible modifications of the URIX model. We have modified the regularization of
the profile functions $Y(r)$ and $T(r)$ at the origin and we have introduced
the $Z_0(r)$ function in the central repulsive $E$-term. We have used one-parameter
functions that have been chosen to match the short-range behavior of the
corresponding functions in the N2LOL model. Furthermore 
the strengths of the $b$-, $d$-terms and $E$-terms have been fixed to reproduce
the triton binding energy and $^2a_{nd}$. Then the predictions for some
selected scattering observables in $p-d$ scattering at 3 MeV have been
compared to the results of the original model and the experimental data.
We can observe that the description using the new parametrizations
has the same quality of the original model. However, with the proposed parametrizations,
the AV18+URIX model describes correctly $B$($^3$H) and $^2a_{nd}$. This analysis can
be consider as a preliminary step in a study directed to determine the parametrizations
of the profile functions inside the three-body force from the experimental data.
Investigations in this direction are underway.

\end{document}